\documentstyle[emulapj]{article}

\newcommand{\beq}{\begin{equation}}
\newcommand{\eeq}{\end{equation}}

\font\tenbg=cmmib10 at 10pt

\def \rvecphi{{\hbox{\tenbg\char'036}}}

\begin{document}
\title{Relativistic Poynting Jets from Accretion Disks}

\author{
R.V.E. Lovelace, and
M.M. Romanova }

\affil{Department of Astronomy,
Cornell University, Ithaca, NY 14853-6801;
RVL1@cornell.edu; Romanova@astro.cornell.edu}

\begin{abstract}

   A model is derived for  relativistic Poynting jets
from the inner region of a disk around a rotating
black hole which is initially
threaded by a dipole-like magnetic field.
    The model is derived from the special
relativistic  equation
for a force-free electromagnetic  field.
    The ``head'' of the Poynting jet is found
to propagate outward with a velocity which
may be relativistic.
   The  Lorentz factor of the head
is $\Gamma =[{B_0^2 / (8 \pi {\cal R}^2 \rho_{ext}c^2}
]^{1/6}$ if this quantity is  much larger than unity.
For conditions pertinent to an active galactic nuclei,
$\Gamma
\approx 8 (10/{\cal R})^{1/3} ({B_0/
10^3 {\rm G}})^{1/3} ({ 1/{\rm cm}^3/
n_{ext}})^{1/6}$, where
$B_0$ is the magnetic field strength close
to the black hole,  $\rho_{ext}=\bar m n_{ext}$ is the
mass density of the ambient medium into which
the jet propagates,  ${\cal R}=r_0/r_g >1$, where
$r_g$ is the gravitational radius of the black hole,
and $r_0$ is the radius of the $O-$point of
the initial dipole field.  This model offers an
explanation for the observed Lorentz factors $\sim 10$
of parsec-scale radio jets measured with very
long baseline interferometry.

\end{abstract}

\section{Introduction}

Highly-collimated,
oppositely directed jets are
observed in
active galaxies and
quasars (see for example
Bridle \& Eilek 1984;
Zensus, Taylor, \& Wrobel 1998),
and in
old compact stars in binaries
(Mirabel \& Rodriguez 1994;
Eikenberry {\it et al.} 1998).
    Further, well collimated
emission line jets are
seen in young stellar
objects (Mundt 1985;  B\"uhrke, Mundt,
\& Ray 1988).
     Recent work favors
models where the twisting of an
ordered magnetic field
threading an accretion
disk acts to magnetically
accelerate the jets (e.g., Meier, Koide,
\& Uchida 2001;  Bisnovatyi-Kogan \& Lovelace 2001).
    There are two regimes:
  (1)  the {\it hydromagnetic regime},
where energy and angular
momentum are carried by both
the electromagnetic field and
the kinetic flux of matter, which
is relevant to the jets from
young stellar objects;
and (2) the {\it Poynting flux regime},
where  energy and angular
from the disk are carried predominantly by the
electromagnetic field, which is relevant
to extra-galactic and microquasar jets,
and possibly to gamma ray burst sources.

   Different theoretical
models have been
proposed for magnetically dominated
or Poynting
outflows (Newman, Newman,
\& Lovelace 1992) and jets
(Lynden-Bell 1996, 2003) from accretion
disks threaded by a dipole-like magnetic field.
    Later,  stationary
Poynting flux dominated
outflows were found in
axisymmetric magnetohydrodynamic
(MHD) simulations of
the opening of  magnetic loops
threading a Keplerian disk (Romanova {\it et al.} 1998).
MHD simulations by Ustyugova {\it et al.} (2000) found
collimated Poynting flux jets.
      The present work represents a continuation
of the studies by Li {\it et al.} (2001)
and Lovelace {\it et al.} (2002; hereafter
L02), which are closely related
to the work by Lynden-Bell (1996).
    Self-consistent ~force-free field solutions
are obtained where the
  twist of each field
line is that due to the differential rotation of a Keplerian disk.
~~

~~

~~

\begin{figure*}[t]
\epsscale{0.8} \plotone{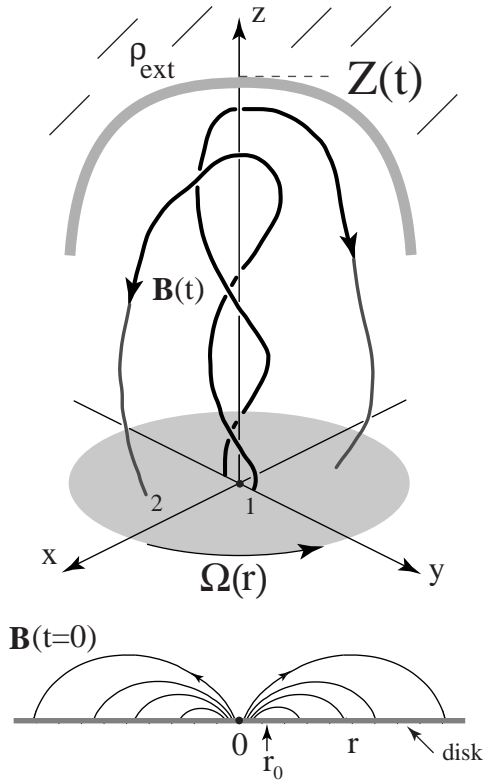} \figcaption{Sketch of the magnetic
field configuration of a Poynting jet.
   The bottom part of the figure shows
the initial  dipole-like magnetic field threading the disk which
rotates at the angular rate $\Omega(r)$.  The top part of the
figure shows the jet at some time later when the head of the jet
is at a distance $Z(t)$.
    At the head of the jet there is
force balance between electromagnetic stress of the jet and the
ram pressure of  the ambient medium of density $\rho_{ext}$.}
\end{figure*}

\section{Theory of Poynting Outflows}

     Here, we consider relativistic
Poynting outflows from a
rotating accretion disk around a Kerr black hole.
    We assume  that at an initial
time $t=0$ an axisymmetric,
dipole-like magnetic field
threads the disk.
    The initial field geometry is
shown in the lower part  of Figure 1.
     This field could result from dynamo processes
in the disk (e.g., Colgate, Li, \& Pariev 1998).
   A more general magnetic field threading
the disk would consist of multiple loops
going from radii say
$r_a$ to $r_b>r_a$ and $r_c>r_b$ to $r_d>r_c$, etc.
It is not clear how to  treat this
case analytically, but non-relativistic MHD
simulations of this type of field show
strong Poynting
flux outflows (Romanova et al. 1998).

     The disk is assumed to be highly conducting
and  dense in the sense that
$\rho_d (r\Omega)^2 \gg {\bf B}^2/4\pi$, $\rho_d$
is the density and $\Omega(r)$ the
angular rotation rate of the disk.
   Thus a magnetic field threading the disk
is frozen into the disk.
     Further, we suppose that
the radial accretion speed of the disk
is much smaller than azimuthal
velocity $\Omega r$.
   For a corotating disk around a Kerr black
hole
\begin{equation}
\Omega ={c^3/(GM) \over a_*+(r/r_g)^{3/2}}~,
\end{equation}
for $r>r_{ms}$ where $r_{ms}$ is the innermost
stable circular orbit, where
$a_*$ is the spin parameter of the black hole
with $0 \geq a_* <1$ and $r_g \equiv GM/c^2$.
For example, for $a_*=0.99$, $r_{ms}\approx 1.45r_g$.
    The simplification is made that
equation (1) also applies for $r<r_{ms}$.
   The region $r \leq r_{ms}$ has a negligible
influence on our results for the considered
conditions where the radial scale of the magnetic
field $r_0$ is such that $(r_g/r_0)^2 \ll 1$.

    In the space above
(and below) the disk we assume a  ``coronal''
or ``force-free'' ideal plasma
(Gold \& Hoyle 1960).
     This plasma may be relativistic with
flow speed ${\bf v}$ comparable to the speed of light.
    Away from the head of the jet, the electromagnetic
field is {\it quasi-stationary}.
   This limit is applicable under conditions
where the energy-density of the plasma $ \gamma \rho c^2$
is much less than the
electromagnetic field energy-density
  $({\bf E}^2+{\bf B}^2)/8\pi$.

    Cylindrical $(r,\phi,z)$
coordinates are used and
axisymmetry is assumed
    Thus the magnetic field has
the form $ {\bf B}~ = {\bf B}_p +
B_\phi \hat{\rvecphi~}~,$ with $
{\bf B}_p = B_r{\hat{\bf r}}+
B_z \hat{\bf z}~.$
We can write
$B_r =-(1 / r)(\partial \Psi/ \partial z),
~B_z =(1 / r)(\partial \Psi / \partial r),$
where $\Psi(r,z) \equiv r A_\phi(r,z)$
is the flux function.
     In the plane of the disk, the flux function
is independent of time owing the frozen-in
condition.
     A representative form of this function is
\begin{equation}
\Psi(r,0)={1\over 2}{r^2B_0 \over 1+2(r/r_0)^3}~,
\end{equation}
where
$B_0$ is the axial magnetic field strength
in the center of the disk, and $r_0$ is
the radius of the $O-$point of the magnetic
field in the plane of the disk as indicated
in Figure 1.
    Equation (2) is taken to apply for
$r \geq 0$ even though it is not valid
near the horizon of the black hole.
As already mentioned the contribution
from region is negligible for the
considered conditions where $(r_g/r_0)^2 \ll 1$.
    Note that $\Psi(r,0)$ has a maximum,
with value $r_0^2B_0/6$, at the
$O-$point where ${\bf B}(r_0,0)=0$.
For  $(r/ r_0)^2 \ll 1$, $\Psi(r,0) \propto r^2$
which corresponds to a uniform vertical field,
whereas for $(r/r_0)^2 \gg 1$, $\Psi(r,0) \propto
1/r$ which corresponds to the dipole field
$B_z(r,0) \propto -1/r^3$.

    It is clear that equations (1) and (2) can
be combined to give $\Omega = \Omega(\Psi)$
which is a double valued function of $\Psi$.
    The upper branch of the function is for the
inner part of the disk ($r \leq r_0$),
while the lower part is for the outer part
of the disk.
     A good  approximation for the upper branch is
obtained by taking
$\Psi \approx r^2 B_0/2$ which
gives $\Omega = (c^3/GM)/[a_*+(2\Psi/B_0r_g^2)^{3/4}]$.

    The main equations for the plasma follow from the
continuity equation ${\nabla \cdot}(\rho{\bf v})=0$,
Amp\`ere's law, ${\bf \nabla \times B}=4\pi {\bf J}/c$,
Coulomb's law ${\bf \nabla \cdot E}= 4\pi \rho_e$,
with $\rho_e$ the charge density, Faraday's law,
${\bf \nabla \times E}=0$, perfect conductivity,
${\bf E}+{\bf v \times B}/c=0$, with ${\bf v}$ the
plasma flow velocity, and the ``force-free'' condition
in the Euler equation, $\rho_e {\bf E}
+{\bf J \times B}/c=0$ (see for example
Lovelace, Wang, \& Sulkanen 1987).
    Owing to the assumed axisymmetry, $E_\phi=0$, so
that the poloidal velocity ${\bf v}_p =
\kappa {\bf B}_p$.
    Mass conservation then gives
${\bf B}\cdot {\bf \nabla} (\rho \kappa) =0,$
which implies that $\rho \kappa = F(\Psi)/4\pi$,
where $F$ is an arbitrary function of $\Psi$.
    In a similar way one finds that
$v_\phi -\kappa B_\phi =r G(\Psi)$, so
that ${\bf E} = - G(\Psi) {\bf \nabla} \Psi$,
and $rB_\phi = H(\Psi)$, so that there are  two
additional functions, $G$ and $H$.

     The function $G$ is determined along
all of the field lines which go through
the disk.
    This follows
from the perfect conductivity condition
at the surface of the disk,
     $z=0$, $E_r +({\bf v \times B})_r/c=0$.
    This gives
$E_r=-(v_\phi B_z - v_z B_\phi)/c =
-v_\phi B_z/c$, where $v_z$ is
zero at the disk and $v_\phi$ is the
disk velocity.
   Therefore,
$E_r(r,0)= -\Omega~ [d\Psi(r,0)/dr]/c$,
so that $G(\Psi)=\Omega(r)/c$ which
gives $\Omega=\Omega(\Psi)$.

   The component of the Euler equation in
the direction of ${\bf \nabla }\Psi$ gives
the force-free
Grad-Shafranov  equation,
\begin{equation}
\left[1-\left({r\Omega \over c}\right)^2\right]\Delta^\star \Psi
-{{\bf \nabla}\Psi \over 2 r^2} \cdot {\bf \nabla}
\left({r^4 \Omega^2 \over c^2 }\right)
+~H {H^\prime}=0~,
\end{equation}
with $ \Delta^\star \equiv
{\partial^2 / \partial r^2}
-(1/r)(\partial / \partial r)
+{\partial^2 / \partial z^2}$ and $H^\prime =
dH(\Psi)/d\Psi$
(Scharlemann \& Wagoner 1973;  Lovelace,
Wang, \& Sulkanen 1987).

      As mentioned above, we consider
an {\it initial value problem} where the
disk at $t=0$ is threaded by a
dipole-like poloidal magnetic field.
      The form of $H(\Psi)$ in
equation (3)
is then determined
by the differential rotation of the
disk:
    The azimuthal {\it twist} of a given field
line going from an inner footpoint
at $r_1$ to an outer footpoint at $r_2$
is fixed by the differential rotation
of the disk.

    Amp\`ere's law gives
$\oint d{\bf l}\cdot {\bf B}
=(4\pi/c)\int d{\bf S}\cdot {\bf J}$,
so that
$rB_\phi(r,z)=H(\Psi)$ is $(2/c)$ times
the current flowing through a circular
area of radius $r$ (with normal $\hat{\bf z}$)
labeled by $\Psi(r,z)$= const.
   Equivalently, $-H[\Psi(r,0)]$
is  $(2/c)$ times
the current flowing into the
area of the disk $\leq r$.
    Our previous work (Li {\it et al.} 2000;
L02) shows that
$-H(\Psi)$ has a maximum so that
the total current flowing into
the disk for $r\leq r_m$ is
$I_{tot} = (c/2)(-H)_{max},$
where $r_m$ is such that
$-H[\Psi(r_m,0)]=(-H)_{max}$
where $r_m$
is less than
the radius of the $O-$point, $r_0$.
   The same total current $I_{tot}$
flows out of the region of the disk
$r=r_m$ to $r_0$.

   For a given field line
we have $rd\phi/B_\phi = ds_p/B_p$,
where $ds_p \equiv \sqrt{dr^2+dz^2}$ is the
poloidal arc length along the field
line, and
$B_p \equiv \sqrt{B_r^2+B_z^2}$.
    The total twist of a field line
loop is
\begin{equation}
\Delta \phi(\Psi) =
-\int_1^2 ds_p ~{B_\phi \over r B_p}
=-H(\Psi) \int_1^2 {ds_p \over r^2 B_p}~,
\end{equation}
with the sign included to give
$\Delta \phi >0$.
    The integration goes from the
disk at a radius $r_1 <r_0$ out into the
corona and back to the disk at a radius
$r_2>r_0$.
   For a prograde disk
the field line twist after a time $t$ is
$
\Delta \phi(\Psi)
=\Omega(r_0)t \{[\Omega(r_1) / \Omega(r_0)]
-[\Omega(r_2)/ \Omega(r_0)]\}.
$

\subsection{Poynting Jets}

     Our previous study of non-relativstic
Poynting jets by analytic theory (L02)
and axisymmetric
  MHD simulations  (Ustyugova et al. 2000) showed
that as the twist, as measured
by  $\Omega(r_0) t$, increases
a new, high twist  field configuration appears
with a different topology.
     A  ``plasmoid''
consisting of toroidal flux
detaches from the disk and propagates
outward.
     The plasmoid
is bounded by a poloidal field
line which has an $X-$point above
the $O-$point on the disk.
   The occurrence of the $X-$point
requires that there be at least
a small amount of dissipation in
the evolution from the poloidal
dipole field and the Poynting jet
configuration.
    The high-twist configuration consists
of a region near the axis which
is {\it magnetically collimated} by
the toroidal $B_\phi$ field and
a region far from the axis
which
is {\it anti-collimated} in the sense
that it is pushed away from the axis.
      The field lines returning
to the disk at $r>r_0$ are
anti-collimated by the pressure
of the toroidal magnetic field.
   The {\it poloidal field} fills only
a small part of the coronal space.

     In the case of relativistic Poynting
jets we hypothesize that the magnetic
field configuration is similar that
in the non-relativistic limit (L02,
Ustyugova et al. 2000).
   Thus, most of the twist
$\Delta \phi$ of a field line of
the relativistic Poynting
jet occurs along the
  jet from $z=0$ to $Z(t)$ as
  sketched in Figure 1, where $Z(t)$ is the
axial location of the
``head'' of the jet.
        Along most of the distance $z=0-Z$ the
radius of the jet is a constant and $\Psi=
\Psi(r)$ for $Z \gg r_0$.
    Note that the function $\Psi(r)$ is different
from $\Psi(r,0)$ which is the flux function
profile on the disk surface.
    Hence,
$r^2 {d\phi/ dz} = {r B_\phi(r,z) / B_z(r,z)}.$
We take for simplicity that $V_z =dZ/dt ={\rm const}$.
   We determine $V_z$ in \S 3.
In this case,
$
   H(\Psi)=[r^2 \Omega(\Psi) / V_z]B_z,
$
where the right-hand side can be written as a function of
$\Psi$ and $d\Psi/dr$.
     This relation allows the closure of equation (3)
which can now be written as
\begin{equation}
{d^2\Psi \over dr^2}+{\lambda -1 \over \lambda +1}
{1\over r}{d\Psi \over dr}+
{\lambda \over \lambda +1}
\left({d\Psi \over dr}\right)^2
{1\over \Omega}{d\Omega \over d\Psi}=0~,
\end{equation}
where $\lambda  \equiv (r\Omega/c)^2[(c/V_z)^2 - 1]=
(r\Omega/c)^2/(\Gamma^2-1)$, and where
$\Gamma \equiv 1/(1-V_z^2/c^2)^{1/2}$.

   Solution of equation (5) is facilitated
by introducing dimensionless variables.
   We measure the radial distance in units of the
distance $r_0$ to the $O-$point of the magnetic
field threading the disk (equation 2).
     We measure the flux function $\Psi$ in
units of $\Psi_0 \equiv r_0^2 B_0/2$.
    The fields are measured in units of $B_0$
which is the magnetic field strength at
the center of the disk.
   Thus, $\bar B_z =(2\bar r)^{-1} d \bar \Psi /d \bar r$.
   The disk rotation rate $\Omega$ is measured
in units of $c^3/(GM)$ (equation 1).
    In terms dimensionless variables $\bar r$,
$\bar \Psi$, and $\bar \Omega$,
equation (5) is the same with overbars on these
three variables.
   Note also that
$\lambda = \bar r^2 {\cal R}^2 \bar \Omega^2/
(\Gamma^2-1)$, where ${\cal R}^2\equiv (r_0/r_g)^2 \gg 1$.

    We consider solutions of eqn. (5)
using the approximation to eqn. (2) of
$\bar \Psi(r,0) = \bar r^2$ which gives
$\bar \Omega =1/(a_*+{\cal R}^{3/2} \bar \Psi^{3/4})$.
     These solutions are close to those
obtained using the full dependence $\Omega(\Psi)$.
    We then have $\bar \Omega =
1/(a_*+{\cal R}^{3/2} \bar \Psi^{3/4})$.
     In this approximation
there is  a unique
self-similar solution to eqn. (5)
for $\bar \Psi \gg 1/{\cal R}^2$
  where   $\bar \Omega \approx
{\cal R}^{-3/2} \bar \Psi^{-3/4}$.
    This solution is
$
\bar \Psi =  {\bar r^{4/3} /
[2 {\cal R}(\Gamma^2 -1)]^{2/3}},
$ with
$
  \lambda =2.
$
    The dependence on $r$ is the same as found
in the non-relativistic case by L02.
   The dependence holds for
$
\bar r_1 ={[2(\Gamma^2-1)]^{1/2} /{\cal R}} <
\bar r < \bar r_2 =
{[2 (\Gamma^2-1)]^{1/2}{\cal R}^{1/2}/ 3^{3/4}}.
$
At the inner radius $\bar r_1$, $\bar \Psi = 1/{\cal R}^2$,
which corresponds to the streamline which passes
through the disk at a distance $r=r_g$.
    For $\bar r < \bar r_1$, we assume
$\bar \Psi \propto \bar r^2$, which
corresponds to $B_z=$ const.
    At the outer radius $\bar r_2$,
$\bar \Psi =(\bar \Psi)_{max} =1/3$ which
corresponds to the streamline which goes
through the disk near the $O-$point at
$r=r_0$.
  Note that there is an appreciable range of radii
if ${\cal R}^{3/2} \gg 1$.
   The non-zero
field components of the Poynting jet are
$
\bar E_r = -\sqrt{2}(\Gamma^2-1)^{1/2} \bar B_z,~
$
$
\bar B_\phi = -\sqrt{2}\Gamma\bar B_z,~
$ and
$
\bar B_z = {2\over 3} (\bar r^{-2/3})
/[2{\cal R}(\Gamma^2-1)]^{2/3},
$
which hold for $r_1 < r < r_2$.
  This electromagnetic field satisfies
the radial force balance equation,
$
{d B_z^2 / dr}+ (1/ r^2)){d [ r^2(B_\phi^2 -E_r^2)]
/ dr} =0,
$
as it should.
    At the outer radius of the jet at $r_2$,
there is a boundary layer where
the axial field changes from $B_z(r_2)$
to zero while (minus) the toroidal field
increases by a corresponding
amount so as to give radial force balance.
   This gives $-r_2B_\phi(r_2)=(-H)_{max}=2I_{max}/c
=(2/\sqrt{3})\Psi(r_2)/r_2$ which
is close the relation found in
the non-relativistic case (L02).

\section{Force Balance at the Head of the Jet}

In the rest frame of the central object
the axial momentum flux-density is $T_{zz}$,
where $T_{jk}$, $(j,k=0,1,2,3)$ is the
usual relativistic
momentum flux-density tensor.
In the reference frame comoving with the front
of the Poynting jet, a Lorentz transformation
gives
$
T_{zz}^\prime = \Gamma^2[T_{zz}
-2(V_z/c)T_{0z} +(V_z/c)^2 T_{00}],
$
where
$
T_{zz}=(E_r^2+B_\phi^2 -B_z^2)/8\pi
=B_z^2(4 \Gamma^2-3)/8\pi,
$
where the last equality uses our
expressions for the fields.
Further,
$
T_{0z} ={E_r B_\phi/ 4 \pi}
={B_z^2}[4\Gamma(\Gamma^2-1)^{1/2}]/8\pi,
$
and
$
T_{00}=(E_r^2 +B_\phi^2+B_z^2)/8\pi
={B_z^2 }(4\Gamma^2-1)/8\pi.
$
Combining these equations gives
$
T_{zz}^\prime = {B_z^2 / 8 \pi}.
$
Since $T_{zz}^\prime$ varies with
radius, we take its average between
$r_1$ and $r_2$ which gives
$
< T_{zz}^\prime >
=B_0^2 /[ 8\pi {\cal R}^2 (\Gamma^2-1)^2].
$
assuming $r_1 \ll r_2$.

   In the reference frame
comoving with the head of the jet,
force balance implies that the
ram pressure of the external
ambient medium $\rho_{ext}\Gamma^2 V_z^2$
is equal to $< T_{zz}^\prime >$.
   This assumes  $V_z$  much larger
than the sound speed in the external medium;
however,
details of the shock(s) at the jet
front are not considered here.
Thus we have
$
{(\Gamma^2-1)^3 }
={B_0^2 /[8 \pi {\cal R}^2 \rho_{ext} c^2]}.
$
For  $\Gamma \gg 1$,
\begin{equation}
\Gamma\approx 8 \left({10 \over {\cal R}}\right)^{1/3} \left({B_0 \over
10^3 {\rm G}}\right)^{1/3} \left({ 1/{\rm cm}^3 \over
n_{ext}}\right)^{1/6}~,
\end{equation}
where $n_{ext} = \rho_{ext}/\bar m$, with $\bar m$
the mean particle mass.
   A necessary condition for the validity of
eqn. (6) is that the axial speed
of the counter-propagating
  fast magnetosonic wave (in the lab
frame) be larger than
$V_z$ so that the jet is effectively
`subsonic.'
   This condition can be expressed as
$\Gamma^2 < v_A^\prime/2$,
where $v_A^\prime
\equiv |{\bf B}^\prime|/(4\pi \rho^\prime)^{1/2}$
is the Alfv\'en speed
in the comoving frame of the jet.
    The above-mentioned force-free condition
is $(v_A^\prime)^2 \gg 1$.
    The other important condition is
that ${\cal R}^2=(r_0/r_g)^2 \gg 1$ so
that the jet energy is extracted mainly
from the disk (rather than from the rotating
black hole).

   The derived value
of $\Gamma$ is of the order of the
Lorentz factors of the expansion
of parsec-scale
extragalactic radio jets observed with
very-long-baseline-interferometry (see,
e.g., Zensus {\it et al.} 1998).
    This interpretation assumes that the
radiating electrons (and/or positrons) are
accelerated to high
Lorentz factors ($\gamma \sim 10^3$) at the jet
front and move
with a bulk Lorentz factor $\Gamma$ relative
to the observer.
     The luminosity of the $+z$ Poynting
jet is $\dot{E}_j=c\int_0^{r_2} r dr E_rB_\phi/2 =
cB_0^2 {\cal R} r_g^2/3 \approx 2.2\times 10^{45}
(B_0/10^3{\rm G})^2 ({\cal R}/10)(M/10^9M_\odot)^2{\rm erg/s}$,
where $M$ is the mass of the black hole.

    The region of collimated field of
the Poynting jet may be
kink unstable.
    The instability will lead to a helical
distortion of the jet with
the helix having the same twist about the
$z-$axis as the axisymmetric ${\bf B}$ field.
    A relativistic perturbation
analysis is  required including the displacement
current.
     The evolution of the
jet evidently depends on {\it both}
the speed of propagation of the lateral
displacement and
the velocity of propagation
of the ``head'' of the jet $V_z$.
    Relativistic propagation of the jet's
head may act to limit the amplitude
of helical kink distortion of the jet.

       For long time-scales,
the Poynting jet is of course time-dependent
due to the angular momentum it extracts
from the inner  disk ($r<r_0$) which causes
$r_0$ to decrease with time (L02).
     This loss of angular momentum leads to
a ``global magnetic instability'' and collapse
of the inner disk (Lovelace {\it et al.}1994, 1997; L02)
and a corresponding outburst of energy in the
jets from the two sides of the disk.
    Such outbursts may explain the flares
of active galactic nuclei blazar sources
(Romanova \& Lovelace 1997; Levinson 1998)
and the one-time outbursts of gamma ray burst
sources (Katz 1997).

     We thank an anonymous referee for
criticism which improved this work and
Hui Li and Stirling Colgate for valuable
discussions.
This work was supported in part by NASA grants
NAG5-9047, NAG5-9735,  by NSF grant AST-9986936,
and by DOE cooperative agreement DE-FC03-02NA00057.
MMR received partial support
from NSF POWRE grant
AST-9973366.

\end{document}